\documentclass[conference]{IEEEtran}

\usepackage[utf8]{inputenc}

\usepackage{cite}

\usepackage[cmex10]{amsmath}
\usepackage{amssymb}

\usepackage{array}

\usepackage{color}

\usepackage{placeins}

\begin{document}

\title{A formal approach for\\ correct-by-construction system
  substitution}

\author{\IEEEauthorblockN{Guillaume Babin\IEEEauthorrefmark{1}\IEEEauthorrefmark{3}}
\IEEEauthorblockA{\IEEEauthorrefmark{1}%
  Université de Toulouse ; INP,  \textit{IRIT} ; 2 rue Camichel, BP 7122, 31071 Toulouse Cedex 7, France}
\IEEEauthorblockA{\IEEEauthorrefmark{3}%
  \textit{CNRS} ; Institut de Recherche en Informatique de Toulouse ; Toulouse, France\\%
\texttt{guillaume.babin@irit.fr}}}

\maketitle

\begin{abstract}
  The substitution of a system with another one may occur in several
  situations like system adaptation, system failure management, system
  resilience, system reconfiguration, etc. It consists in replacing a
  running system by another one when given conditions hold. This
  contribution summarizes our proposal to define a formal setting for
  proving the correctness of system substitution. It relies on
  refinement and on the Event-B method.
\end{abstract}

\vspace{2mm}

\section{Motivation}

Several efforts were devoted to the formal development of complex
systems. Different formal approaches have been defined, they led to
numerous system developments in real life applications like
transportation systems, web applications, information and data
management, etc.

One of the major concerns in system development relates to the
substitutability of a system by another one. This capability can be
seen throughout different concerns. Among these ones we are mainly
interested in two of them: functional and behavioural concerns.
\begin{enumerate}
\item The \textit{functional concern} deals with the function of the
  system i.e. what it does. We have identified three main
  substitutions:
  \begin{itemize}
  \item[-] a system may be replaced by an equivalent one;
  \item[-] a system may be degraded i.e. replaced by a system
    achieving less functions, the problem being to identify that the
    critical ones are still achieved;
  \item[-] a system may be upgraded i.e. replaced by a system
    achieving more functions.
  \end{itemize}
  The functional concern relates to the preservation of the functions
  of the substituted system by the substitute one. This preservation
  is studied through property verification. Classical formal
  verification techniques guarantee that the substituted and the
  substitute systems refine the same specification.
\item The \textit{behavioural concern} deals with the behaviour of a
  system in terms of execution sequences. We have identified two main
  substitutions:
  \begin{itemize}
  \item[-] a system may be replaced by another system, but it requires
    to restart the system from its initial state. This corresponds to
    a \textit{cold start} substitution;
  \item[-] a system may be replaced at a given state by another system
    recovering this state and pursuing the execution of the system
    from this recovered state. This corresponds to a \textit{warm
      start} or \textit{hot start} substitution.
  \end{itemize}
  The behavioural concern requires an explicit modelling of system
  states during execution.
\end{enumerate}
Through these concerns several properties like dependability
\cite{Laprie1995}, adaptation \cite{Weyns2012} \cite{Kell2008},
resilience, self-healing systems \cite{Parashar2005}, loss of QoS can
be addressed.

System substitution is performed when an event requiring this change
occurs. In several cases, this event is (or these events are)
triggered by monitors in case of monitored systems, or by evaluating a
given situation in autonomous systems (\mbox{self-$\star$} systems).

This paper overviews the formal approach we propose to handle system
substitutions.

\section{Our approach}

In the following, we consider systems that are expressed as states and
transitions describing state changes. Our approach is a
correct-by-construction one, it relies on refinement and on the
Event-B method \cite{Abrial2010} \cite{Abrial2007}
\cite{Ait-Ameur2005} \cite{Ait-Ameur2009}. It addresses system
substitution and covers both concerns, functional and behavioural.

\subsection{The functional concern: refinement}

Refinement relates an abstract system to a concrete one. The refined
concrete system preserves the properties of the abstract one. Both of
the three identified cases of the functional concern may be encoded by
refinement. Indeed, several concrete representations of a given
specification can be given at different abstraction levels, linked by
a refinement relationship. Invariants represent the key feature for
describing equivalent, degraded or upgraded systems.

\subsection{The behavioural concern: refinement and variants}

Once the functional concern has been addressed, it becomes possible to
assert that a system may substitute another one in one of the three
identified cases above. When a system runs, one obvious substitution
case corresponds to a re-start of the substitute system. But, when the
sequence of system running states is recorded, one may identify
correlations between the states of the substitute system and the
current state of the substituted system. Our proposal is to use
explicit variants to identify the substitution state and invariants to
express such state correlation. Indeed, variants describe the sequence
of running states. Correspondences between states of the substituted
and of the substitute system are expressed by the variant
values. Invariants define properties between the states of both
systems, enabling the copy of the substituted state variables to the
substitute state variables.

\subsection{Why Event-B ?}

The Event-B method is set-up in order to formalise our
approach. Event-B is a state-based method that promotes
correct-by-construction development paradigm and formal verification
by theorem proving. Indeed, Event-B supports the definition of
state-based systems where transitions recording state changes are
encoded by events. Moreover, it offers the capability to explicitly
express variants and invariants, and to build systems using
refinements.

\subsubsection{Event-B refinement for the functional concern} a top
level machine, representing the main function(s) of a system may be
refined by one or more other machines leading to different concrete
system designs. All the obtained machines are possible system
substitutes.

\subsubsection{Event-B refinement and variants for the behavioural
  concern} it is possible to link two substitute systems of the same
system into a single refinement. The idea consists in linking these
two substitutes with a property establishing a relation between the
state variables of one substitute system and the
\textit{corresponding} state variables of the other substitute
system. We call this property an \textit{horizontal invariant}. The
corresponding state is defined thanks to explicit variants. Indeed, a
state of the substitute system can recover the state of another
substitute system if their variant values match. A specific event
\textit{switch} commutes to the substitute system in a single Event-B
model of both systems.

\section{Formalisation with Event-B}

\subsection{Functional concern}

The first step consists in defining a top level specification Event-B
machine. Each system, refining this specification is considered as a
potential substitute. Obviously, the chosen invariant defines the
nature of this substitution (equivalence, degradation or
upgrade). Once these refinements have been conducted, we get a set of
potential substitutes.

\subsection{Behavioural concern}

The behavioural concern requires the explicit manipulation of the
current state of the system to be substituted. Therefore, our model
defines systems as pairs $(sv,VarExp)$:
\begin{itemize}
\item[-] $sv$ being a subset of $Vars$ the set of system variables,
  such that $sv \subseteq Vars$
\item[-] an expression $VarExp=exp(v_1,\cdots,v_n)$ over $v_i \in sv$
  evaluating to a natural number and defining a decreasing
  function. This expression describes a variant for the studied
  system.
\end{itemize}

This model, combining functional and behavioural concerns, is generic
enough to enable the expression of a system together with an explicit
representation of its states, if combined with a valuation of its
variables. This explicit representation makes it possible to
explicitly manipulate a state in an Event-B machine. As a consequence
of this modelling, the expression of generic mechanisms combining
variant values for recovering states becomes possible.

\subsection{Fragment of the Event-B generic model}

The following Event-B fragment results from the previously described
model.
\begin{itemize}
\item[-] $Variables$ defines the carrier set containing all the system
  variables
\item[-] $Values$ is the carrier set of the values of the variables
\item[-] $VariablesSets \subseteq \mathcal P \left({Variables}\right)$
  a set of sets of variables to identify the variables of each system
  (partition)
\item[-] $Valuations = Variables \rightarrow Values$ functions that
  give the value of a variable
\item[-] $Systems = VariablesSets \times (Valuations \rightarrow
  \mathbb{N})$ a system is a pair composed of a group of variables and
  a variant function
\item[-] $Systems\_states = Systems \times Valuations$: all the states
  of the systems, i.e. pairs (system -- variables values)
\end{itemize}

\section{Case study}
To illustrate our approach a case study issued from electronic
commerce is shown.  We consider an online purchase of goods composed
of four steps: \textit{selection} of goods by filling a cart,
\textit{payment}, \textit{billing} and \textit{delivery}. Solely the
\textit{selection} step is detailed below. Moreover, for this case
study, we consider that the event triggering the system substitution
is a system failure. A failure may occur while the client is filling
his cart with goods during the \textit{selection} step. Failure is the
event that triggers system substitution.

Setting up our approach led to the following steps.

\subsection{Functional concern}
\begin{itemize}
\item[-] an upper model level with an abstract selection of a set of
  goods in a cart has been designed (corresponding to the Event-B
  machine \textit{M1});
\item[-] a first basic selection system (\textit{Sys1}) composed of
  one cart located on a website, has been created by refining
  \textit{M1} (machine \textit{M11});
\item[-] a second basic selection system (\textit{Sys2}) composed of
  two carts located on different websites, without failures, has been
  created by refining \textit{M1} (machine \textit{M12});
\item[-] a system composed of the previous ones. The used system is
  chosen at initialisation (machine \textit{M13} refining
  \textit{M1}).
\end{itemize}

\subsection{Behavioural concern}

\begin{itemize}
\item[-] The first step towards handling the behavioural concern was
  the creation of an abstract selection feature, refining \textit{M1}
  and introducing the possibility to switch from \textit{Sys1} to its
  substitute \textit{Sys2} . Here, we detail the switching process
  only. (Machine \textit{M14})
\item[-] {\it Cold start. } \textit{M14} is refined, by defining a
  mechanism that substitutes \textit{Sys1} by \textit{Sys2} with a
  reinitialisation on the initial state of \textit{Sys2}.  The event
  triggering the substitution re-initialises \textit{Sys2} from its
  initial state. (Machine \textit{M14})
\item[-] {\it Hot or Warm start. } \textit{M14} is refined to a
  machine where \textit{Sys1} is substituted by \textit{Sys2} with
  preservation of the previous \textit{Sys1} executions. The event
  that triggers the substitution restores the current state of
  \textit{Sys2} to a state that functionally matches with
  \textit{Sys1} state before substitution. By functionally, we mean
  that there is no selected goods of \textit{Sys1} cart that are
  lost. (Machine \textit{M142})
\end{itemize}


\begin{table}
  \centering
  \caption{Proofs statistics}
  \label{stats}
  \begin{tabular}{ | l | c | c | c c |}
    \hline
    Event-B  & Total & Automatic & \multicolumn{2}{|c|}{Interactive} \\
    Machine  & PO    & proof     & \multicolumn{2}{|c|}{proof} \\ \hline
    \hline
    \textit{M1}   & 31    & 27   & 4  & (13\%) \\ \hline
    \textit{M11}  & 28    & 27   & 1  & (4\%)  \\ \hline
    \textit{M12}  & 57    & 56   & 1  & (2\%)  \\ \hline
    \textit{M13}  & 99    & 97   & 2  & (2\%)  \\ \hline
    \textit{M141} & 133   & 129  & 4  & (3\%)  \\ \hline
    \textit{M142} & 230   & 214  & 16 & (7\%)  \\ \hline
    \hline
    Generic model & 37    & 28   & 9  & (24\%) \\ \hline
    Instanciation & 53    & 39   & 14 & (26\%) \\
    \hline
  \end{tabular}
\end{table}


\subsection{Instanciation of the generic model}

We instantiated our model, according to the previously overviewed
generic model, with the following definitions.
\begin{itemize}
\item[-] $Variables = \{C1, C2a, C2b\}$ corresponding to the
carts. $C1$ for \textit{Sys1} and $C2a, C2b$ for \textit{Sys2};

\item[-] $ValueElements = \{Prod1, Prod2, Prod3, Prod4, \\
\hspace*{1cm} Prod5\}$ is the set of goods;

\item[-] $Valuations = (\{C1\} \mapsto \mathcal P(ValueElements)) \\
\hspace*{1cm} \cup (\{C2a,C2b\} \mapsto \mathcal P(ValueElements))$
associates  a subset of goods to each cart;

\item[-] $VariablesSets = \{\{C1\},\{C2a,C2b\}\}$
identifies the variables of each system \textit{Sys1} and \textit{Sys2};

\item[-] $Sys1 = \{C1\} \mapsto (\lambda val \cdot val \in \\
\hspace*{1cm} \{C1\} \rightarrow \mathcal P(ValueElements) \mid \\
\hspace*{2cm} card(ValueElements)-card(val(C1)))$

\item[-] $Sys2 = \{C2a,C2b\} \mapsto (\lambda val \cdot val \in \{C2a,C2b\} \rightarrow \\
\hspace*{1cm} \mathcal P(ValueElements) \mid card(ValueElements) \\
\hspace*{2cm} - card(val(C2a) \cup val(C2b)))$

\item[-] $Systems = \{Sys1,Sys2\}$

\item[-] $Systems\_states = Systems \times Valuations$
\end{itemize}

We were then able to refine this machine with one expressing directly
$C1$, $C2a$ and $C2b$ carts, and prove the refinement. Moreover, the
following safety properties have been proved on each of the machines.
\begin{itemize}
\item[-] All the desired products are selected after the
  \textit{selection} action. If $P \subseteq PRODUCTS$ is the set of
  products to purchase, and $carts \in SITES \times PRODUCTS$ is the
  variable denoting the pair of selected products on a given
  website. This property is expressed as
  \begin{center}
    $selection\_done \Rightarrow \text{ran}(carts) = P$
  \end{center}

\item[-] There is no product selected twice i.e. no product in both of
  the two carts, expressed as
  \begin{center}
    $\forall p, \, p \in \text{ran}(carts) \Rightarrow
    \text{card}(carts^{-1}[\{p\}]) = 1$
  \end{center}
\end{itemize}

Here, the horizontal invariant is $val(C1) = val(C2a) \cup val(C2b)$,
where $val$ is the corresponding valuation.

This generic model together with this case study have been modelled on
the RODIN platform \cite{Abrial2006} \cite{Abrial2010a} and the
statistics of table \ref{stats} (proof obligations PO) have been
obtained.

\section{Ongoing work}

This paper presented a global view of our approach for system
substitution. This approach exploits the notion of refinement,
invariants and variants. System substitution has been illustrated with
a use case from electronic commerce.

Currently, our work is pursued in two directions. On the one hand, we
are building a generic model for system substitution formalised within
the Event-B method. A set of generic machines, to be instantiated,
defining system substitution is under construction. It is planned to
study within this framework adaptation, self-healing, reliability,
\ldots situations. On the other hand, we plan to integrate a monitor
in order to monitor a property of the system behaviour and trigger
system substitution. For example, monitored properties could be
identification of system failures or loss of quality of service.

\bibliographystyle{IEEEtran}
\bibliography{IEEEabrv,database}

\end{document}